Original article

Corresponding Author:

LIU Jiangyi , Electronic and optical engineering Department, Shijiazhuang Campus of Army Engineering University，Shijiazhuang, 050003, China.

Email: liu_jiangyi@163.com

# Particle Probability Hypothesis Density Filter based on Pairwise Markov Chains

LIU Jiangyi[1] WANG Chunping[1] WANG Wei[2]

1. Electronic and optical engineering Department, Shijiazhuang Campus of Army Engineering University，Shijiazhuang, China；2. China Huayin Ordnance Test Center, Huayin, Shaanxi, China

## Abstract

Most multi-target tracking filters assume that one target and its observation follow a Hidden Markov Chain (HMC) model, but the implicit independence assumption of HMC model is invalid in many practical applications, and a Pairwise Markov Chain (PMC) model is more universally suitable than traditional HMC model. A particle probability hypothesis density filter based on PMC model (PF-PMC-PHD) is proposed for the nonlinear multi-target tracking system. Simulation results show the effectiveness of PF-PMC-PHD filter, and that the tracking performance of PF-PMC-PHD filter is superior to the particle PHD filter based on HMC model in a scenario where we kept the local physical properties of nonlinear and Gaussian HMC models while relaxing their independence assumption.

## 1 Introduction

Random Finite Set (RFS) theory has been widely used in multi-target tracking field. Unlike traditional solutions based on data association, RFS-based solutions provide a theoretical framework without data association [**1,2**]. Among RFS based solutions, the Probability Hypothesis Density (PHD) filter propagates the first order moment of the posterior multi-target density[**3**], which is now widely applied . RFS based solutions can not get the analytical solution directly, and the implementations mainly based on numerical approximations, such as Gauss Mixture (GM) PHD filter [**4,5**], and on Sequential Monte Carlo (SMC, i.e. particle filter) methods[**6,7**].

Most multi-target tracking filters, including the classical PHD filter, assume that the targets and the observations they produce follow the well known HMC model. HMC model assumes that the state of a given target is a Markov Chain (MC), however, the Markovian and independence assumption implicit in the HMC model may not be invalid in practical applications [**8**]. In 2000, Pieczynski proposed the PMC model in order to relax the independence assumption of the HMC model [**9-11**], HMC model is a special PMC model, and PMC model is more universally suitable than HMC model. In 2013, Petetin and Desbouvries proposed a PHD filter for targets follow the PMC model(PMC-PHD) [**8**], and prove that the tracking performance of proposed PMC-PHD filter is better than the "classical" PHD filter based on the HMC model under the relaxed independence assumptions. and the PMC-PHD filter proposed by Petetin only considers the first order information of the target state, neglecting its high order information, and leads to the instability of the target number estimation. In view of this problem, Mahler proposed a Cardinalized Probability Hypothesis Density filter based on PMC model (PMC-CPHD), which is developed from PMC-PHD filter by propagating cardinality distribution function of the target simultaneously[**12**].

The GM implementation of PMC-PHD filter proposed by Petetin and Desbouvries is only suitable for the linear Gaussian multi-target tracking system. In this article, A particle implementation of PMC-PHD (PF-PMC-PHD) filter is proposed for the nonlinear multi-target tracking system based on PMC model. The sampling importance density function of the proposed PF-PMC-PHD filter does not contain the latest measurement information, which may lead to problems such as filtering divergence and particle degradation. Simulation result verifies the effectiveness of the PF-PMC-PHD filter, and shows that the performance of PF-PMC-PHD is better than the particle implementation of the typical HMC-PHD filter (PF-HMC-PHD) in a scenario where we kept the local physical properties of nonlinear and Gaussian HMC models while relaxing their independence assumption.

## 2 PHD filter based on PMC model

### 2.1 PMC model

Let $\boldsymbol{x}_k \in \mathbb{R}^m$ express the state at time $k$, and the corresponding observation is $\boldsymbol{y}_k \in \mathbb{R}^q$, if the joint probability density function(pdf) of $(\boldsymbol{x}_{0:k}, \boldsymbol{y}_{0:k})$ can be factorized as follows:

$$p(\boldsymbol{x}_{0:k}, \boldsymbol{y}_{0:k}) = p(\boldsymbol{x}_0, \boldsymbol{y}_0) \prod_{i=1}^{k} p_{i|i-1}(\boldsymbol{x}_i, \boldsymbol{y}_i \,|\, \boldsymbol{x}_{i-1}, \boldsymbol{y}_{i-1}) \tag{1}$$

where $p(\boldsymbol{x}_0, \boldsymbol{y}_0)$ is the state distribution at the initial time, and the following formula (2) can be true[**13**]:

$$f(\boldsymbol{x}_k \,|\, \boldsymbol{x}_{k-1}, \boldsymbol{y}_{k-1}) \neq f(\boldsymbol{x}_k \,|\, \boldsymbol{x}_{k-1}), \qquad f(\boldsymbol{y}_k \,|\, \boldsymbol{x}_k, \boldsymbol{x}_{k-1}, \boldsymbol{y}_{k-1}) \neq g(\boldsymbol{y}_k \,|\, \boldsymbol{x}_k) \tag{2}$$

where $f(\boldsymbol{x}_k \,|\, \boldsymbol{x}_{k-1})$ and $g(\boldsymbol{y}_k \,|\, \boldsymbol{x}_k)$ are the target Markov transition density and the sensor likelihood function.

Define target motion model follows the PMC model as:

$$\left(\boldsymbol{x}_k, \boldsymbol{y}_k\right) = \varphi\left(\left(\boldsymbol{x}_{k-1}, \boldsymbol{y}_{k-1}\right), \mathbf{w}_k\right) \tag{3}$$

where $\mathbf{w}_1, \ldots, \mathbf{w}_k$ are independent zero-mean Gaussian noises, and

$$E(\mathbf{w}_k \mathbf{w}_k^T) = \Sigma_k = \begin{bmatrix} \Sigma_k^{11} & {\Sigma_k^{21}}^T \\ \Sigma_k^{21} & \Sigma_k^{22} \end{bmatrix} \tag{4}$$

A classical example of Gaussian PMC model can be expressed as[**8**]:

$$\begin{bmatrix} \boldsymbol{x}_k \\ \boldsymbol{y}_k \end{bmatrix} = \underbrace{\begin{bmatrix} \mathbf{F}_k^1 & \mathbf{F}_k^2 \\ \mathbf{H}_k^1 & \mathbf{H}_k^2 \end{bmatrix}}_{\mathbf{B}_k} \begin{bmatrix} \boldsymbol{x}_{k-1} \\ \boldsymbol{y}_{k-1} \end{bmatrix} + \mathbf{w}_k \tag{5}$$

## 2.2 PHD filter based on PMC model

RFS-based solutions consider that at time $k$ targets and measurements are two RFS: $X_k = \{\boldsymbol{x}_1, \ldots, \boldsymbol{x}_n\}$ and $Z_k = \{\boldsymbol{z}_1, \ldots, \boldsymbol{z}_m\}$, where $n$ and $m$ are random integers, which indicate number of targets and measurements. The measurement associated to a given state $\boldsymbol{x}_i$ is noted $\boldsymbol{y}_i$, and define random finite set $\dot{X}_k = \{(\boldsymbol{x}_1, \boldsymbol{y}_1), \ldots, (\boldsymbol{x}_n, \boldsymbol{y}_n)\}$ 。

PHD $v(\boldsymbol{x})$ of random finite set $X$ is the first order moment of multi-target density. Directly computing $v_k(\boldsymbol{x})$ in a PMC seems complicated because $\boldsymbol{x}_k$ is not necessarily Markovian; However, we can propagate a joint intensity $v_k(\boldsymbol{x}, \boldsymbol{y})$, and then obtain the PHD $v_k(\boldsymbol{x})$ as $v_k(\boldsymbol{x}) = \int v_k(\boldsymbol{x}, \boldsymbol{y}) \mathrm{d}\boldsymbol{y}$.

Assuming that there is no spawning (if there is spawning the extension is immediate), according to the multi-target Bayesian principle, the joint PHD can be propagated through the following prediction and update formula:

$$v_{k|k-1}(\boldsymbol{x}, \boldsymbol{y}) = \int p_{S,k}(\boldsymbol{x}_{k-1}) f_{k|k-1}(\boldsymbol{x}, \boldsymbol{y} \mid \boldsymbol{x}_{k-1}, \boldsymbol{y}_{k-1}) \times v_{k-1}(\boldsymbol{x}_{k-1}, \boldsymbol{y}_{k-1}) d\boldsymbol{x}_{k-1} d\boldsymbol{y}_{k-1} + b_k(\boldsymbol{x}, \boldsymbol{y}) \tag{6}$$

$$v_k(\boldsymbol{x}, \boldsymbol{y}) = \left[1 - p_{D,k}(\boldsymbol{x})\right] v_{k|k-1}(\boldsymbol{x}, \boldsymbol{y}) + \sum_{\boldsymbol{z} \in Z_k} \frac{p_{D,k}(\boldsymbol{x}) v_{k|k-1}(\boldsymbol{x}, \boldsymbol{z}) \delta_{\boldsymbol{z}}(\boldsymbol{y})}{\kappa_k(\boldsymbol{z}) + \int p_{D,k}(\boldsymbol{x}) v_{k|k-1}(\boldsymbol{x}, \boldsymbol{z}) d\boldsymbol{x}} \tag{7}$$

where $p_{S,k}(\boldsymbol{x})$ is the probability that a target with state $\boldsymbol{x}$ at time $k$-1 still exists at time $k$; $p_{D,k}(\boldsymbol{x})$ is the probability that a target with state $\boldsymbol{x}$ is detected at time $k$; $b_k(\boldsymbol{x}, \boldsymbol{y})$ is the joint PHD of the birth targets RFS at time $k$; $\kappa_k(\boldsymbol{z})$ is the PHD of the clutter measurements RFS at time $k$; $\delta_{\boldsymbol{z}}(\boldsymbol{y})$ is the Dirac delta function concentrated at $\boldsymbol{z}$ 。 Finally remember that $v_k(\boldsymbol{x}) = \int v_k(\boldsymbol{x}, \boldsymbol{y}) d\boldsymbol{y}$, where the integral w.r.t. $\boldsymbol{y}$ can reduce to a sum.

# 3 PF-PMC-PHD filter

There are two main methods to implement PHD filter, one is particle implementation and the other is GM implementation. The GM implementation of PMC-PHD filter proposed by Petetin and Desbouvries is only suitable for linear Gaussian multi-target tracking system. The particle PMC-PHD (PF-PMC-PHD) is given for the nonlinear multi-target tracking system based on PMC model in this paper.

A set of weighted random samples $\left\{w_k^{(i)},(\boldsymbol{x}_k^{(i)},\boldsymbol{y}_k^{(i)})\right\}_{i=1}^{L}$ are used to approximate the posterior probability density function of the pair $(\boldsymbol{x},\boldsymbol{y})$ as follows:

$$v_k(\boldsymbol{x},\boldsymbol{y}) \approx \sum_{i=1}^{L} w_k^{(i)}\delta((\boldsymbol{x},\boldsymbol{y})-(\boldsymbol{x}_k^{(i)},\boldsymbol{y}_k^{(i)})) \tag{8}$$

where $w_k^{(i)}$ represents the expected value of pair whose state is $(\boldsymbol{x}_k^{(i)},\boldsymbol{y}_k^{(i)})$.

PF-PMC-PHD filter can be summarized as follow.

**Step 1:** Initialization of particles.

At time $k=0$, use $L_0$ particles $\left\{w_0^{(i)},(\boldsymbol{x}_0^{(i)},\boldsymbol{y}_0^{(i)})\right\}_{i=1}^{L_0}$ to represent the prior probability density $v_0(\cdot)$ of pair $(\boldsymbol{x},\boldsymbol{y})$, and particles number is proportional to the number of targets, that is, if there are $\hat{N}_0$ targets, $w_0^{(i)}=\hat{N}_0/L_0$ . joint PHD function $v_0(\boldsymbol{x},\boldsymbol{y})$ of pair $(\boldsymbol{x},\boldsymbol{y})$ writes as:

$$v_0(\boldsymbol{x},\boldsymbol{y}) = \sum_{i=1}^{L_0} w_0^{(i)}\delta((\boldsymbol{x},\boldsymbol{y})-(\boldsymbol{x}_0^{(i)},\boldsymbol{y}_0^{(i)})) \tag{9}$$

**Step 2:** Particles prediction.

At time $k\geq 1$, the particles $(\tilde{\boldsymbol{x}}_k^{(i)},\tilde{\boldsymbol{y}}_k^{(i)})$ generating for surviving targets are sampled from the proposed importance probability density function $q_k(\cdot|(\tilde{\boldsymbol{x}}_{k-1}^{(i)},\tilde{\boldsymbol{y}}_{k-1}^{(i)}),\mathrm{Z}_k)$, $i=1,\ldots L_{k-1}$, and the particles $(\tilde{\boldsymbol{x}}_k^{(i)},\tilde{\boldsymbol{y}}_k^{(i)})$, $i=L_{k-1}+1,\ldots L_{k-1}+J_k$ generating for the new birth targets are sampled from another suggested density function $p_k(\cdot|Z_k)$. The corresponding predicted weights are calculated as:

$$\tilde{w}_{k|k-1}^{(i)} = \begin{cases} w_{k-1}^{(i)}\cdot\dfrac{\phi_{k|k-1}\left((\tilde{\boldsymbol{x}}_k^{(i)},\tilde{\boldsymbol{y}}_k^{(i)}),(\boldsymbol{x}_{k-1}^{(i)},\boldsymbol{y}_{k-1}^{(i)})\right)}{q_k((\tilde{\boldsymbol{x}}_k^{(i)},\tilde{\boldsymbol{y}}_k^{(i)})|(\boldsymbol{x}_{k-1}^{(i)},\boldsymbol{y}_{k-1}^{(i)}),Z_k)} & i=1,\ldots L_{k-1} \\ \dfrac{1}{J_k}\cdot\dfrac{b_k(\tilde{\boldsymbol{x}}_k^{(i)},\tilde{\boldsymbol{y}}_k^{(i)})}{p_k((\tilde{\boldsymbol{x}}_k^{(i)},\tilde{\boldsymbol{y}}_k^{(i)})|Z_k)} & i=L_{k-1}+1,\ldots L_{k-1}+J_k \end{cases} \tag{10}$$

Assuming that there is no spawning,

$$\phi_{k|k-1}\left((\tilde{\boldsymbol{x}}_k^{(i)}, \tilde{\boldsymbol{y}}_k^{(i)}), (\boldsymbol{x}_{k-1}^{(i)}, \boldsymbol{y}_{k-1}^{(i)})\right) = p_{S,k}(\boldsymbol{x}_{k-1}^{(i)})\, f_{k|k-1}(\tilde{\boldsymbol{x}}_k^{(i)}, \tilde{\boldsymbol{y}}_k^{(i)} \mid \boldsymbol{x}_{k-1}^{(i)}, \boldsymbol{y}_{k-1}^{(i)}) \tag{11}$$

Predicted joint PHD function $v_{k|k-1}(\boldsymbol{x}, \boldsymbol{y})$ of pair $(\boldsymbol{x}, \boldsymbol{y})$ writes as：

$$v_{k|k-1}(\boldsymbol{x}, \boldsymbol{y}) = \sum_{i=1}^{L_{k-1}+J_k} \tilde{w}_{k|k-1}^{(i)} \cdot \delta((\boldsymbol{x}, \boldsymbol{y}) - (\tilde{\boldsymbol{x}}_k^{(i)}, \tilde{\boldsymbol{y}}_k^{(i)})) \tag{12}$$

**Step 3:** Particles update.

Recalculating weights of particles using measurements $\boldsymbol{z} \in Z_k$ get from sensor, and the posterior probability density function $v_k(\boldsymbol{x}, \boldsymbol{y})$ of pair $(\boldsymbol{x}, \boldsymbol{y})$ writes as：

$$v_k(\boldsymbol{x}, \boldsymbol{y}) = v_k^1(\boldsymbol{x}, \boldsymbol{y}) + v_k^2(\boldsymbol{x}, \boldsymbol{y}) \tag{13}$$

$$v_k^1(\boldsymbol{x}, \boldsymbol{y}) = \sum_{i=1}^{L_{k-1}+J_k} \tilde{w}_k^{1,(i)} \delta((\boldsymbol{x}, \boldsymbol{y}) - (\tilde{\boldsymbol{x}}_k^{(i)}, \tilde{\boldsymbol{y}}_k^{(i)})) \tag{14}$$

$$\tilde{w}_k^{1,(i)} = \left(1 - p_{D,k}(\tilde{\boldsymbol{x}}_k^{(i)})\right) \tilde{w}_{k|k-1}^{(i)} \tag{15}$$

$$v_k^2(\boldsymbol{x}, \boldsymbol{y}) = \sum_{\boldsymbol{z} \in Z_k} \sum_{i=1}^{L_{k-1}+J_k} \tilde{w}_k^{2,(i)}(\boldsymbol{z}) \delta(\boldsymbol{x} - \tilde{\boldsymbol{x}}_k^{(i)}) \delta_{\boldsymbol{z}}(\boldsymbol{y}) \tag{16}$$

$$\tilde{w}_k^{2,(i)}(\boldsymbol{z}) = \frac{p_{D,k}(\tilde{\boldsymbol{x}}_k^{(i)}) q_k^{(i)}(\boldsymbol{z}) \cdot \tilde{w}_{k|k-1}^{(i)}}{\kappa_k(\boldsymbol{z}) + \sum_{i=1}^{L_{k-1}+J_k} p_{D,k}(\tilde{\boldsymbol{x}}_k^{(i)}) q_k^{(i)}(\boldsymbol{z}) \tilde{w}_{k|k-1}^{(i)}} \tag{17}$$

$$q_k^{(i)}(\boldsymbol{z}) = \mathrm{N}(\boldsymbol{z}; \tilde{\boldsymbol{y}}_{k|k-1}^{(i)}; \Sigma_k^{22}) \tag{18}$$

**Step 4:** Resampling of the particles.

Estimating the number of targets:

$$\hat{N}_k = \sum_{i=1}^{L_{k-1}+J_k} \tilde{w}_k^{1,(i)} + \sum_{\boldsymbol{z} \in Z_k} \sum_{i=1}^{L_{k-1}+J_k} \tilde{w}_k^{2,(i)}(\boldsymbol{z}) \tag{19}$$

Resampling particles $\left\{\tilde{w}_k^{1,(i)} / \hat{N}_k, (\tilde{\boldsymbol{x}}_k^{1,(i)}, \tilde{\boldsymbol{y}}_k^{1,(i)})\right\}_{i=1}^{L_{k-1}+J_k} \cup \left\{\left\{\tilde{w}_k^{2,(i)}(\boldsymbol{z}_j) / \hat{N}_k, (\tilde{\boldsymbol{x}}_k^{2,(i)}, \boldsymbol{z}_j)\right\}_{i=1}^{L_{k-1}+J_k}\right\}_{j=1}^{|Z_k|}$, meanwhile, keeping the value $\boldsymbol{y}$ of particles which represent $v_k^2(\boldsymbol{x}, \boldsymbol{y})$， we will obtain particles $\left\{w_k^{(i)} / \hat{N}_k, (\boldsymbol{x}_k^{(i)}, \boldsymbol{y}_k^{(i)})\right\}_{i=1}^{L_k}$.

**Step 5:** Approximation of posterior probability density.

Rewrites posterior joint PHD $v_k(\boldsymbol{x}, \boldsymbol{y})$ of pair $(\boldsymbol{x}, \boldsymbol{y})$ as：

$$v_k(\boldsymbol{x},\boldsymbol{y})=\sum_{i=1}^{L_k} w_k^{(i)}\delta((\boldsymbol{x},\boldsymbol{y})-(\boldsymbol{x}_k^{(i)},\boldsymbol{y}_k^{(i)})) \tag{20}$$

According to $v_k(\boldsymbol{x})=\int v_k(\boldsymbol{x},\boldsymbol{y})d\boldsymbol{y}$, and $\int \delta(\boldsymbol{y}-\boldsymbol{y}_k^{(i)})d\boldsymbol{y}=1$, $i=1,\ldots L_{k-1}+J_k$, posterior PHD $v_k(\boldsymbol{x})$ of state $\boldsymbol{x}$ writes as:

$$v_k(\boldsymbol{x})=\sum_{i=1}^{L_k} w_k^{(i)}\delta(\boldsymbol{x}-\boldsymbol{x}_k^{(i)}) \tag{21}$$

# 4 Experimental simulation

## 4.1 A particular class of Gaussian PMC model

In order to verify the effectiveness of the proposed PF-PMC-PHD filter, and compare the tracking performance with traditional PF-HMC-PHD filter, the experimental simulation uses a special Gaussian PMC model, whose $p(\mathbf{x}_k|\mathbf{x}_{k-1})$ and $p(\mathbf{y}_k|\mathbf{x}_k)$ are the same as $f_{k|k-1}(\mathbf{x}_k|\mathbf{x}_{k-1})$ and $g_k(\mathbf{y}_k|\mathbf{x}_k)$ of HMC model while relaxing the independence assumption. The two models have the same local physical properties. Suppose that for all $k$, a HMC model satisfies:

$$p(\mathrm{x}_0)=\mathrm{N}(\mathrm{x}_0;\mathrm{m}_0;\mathrm{P}_0) \tag{22}$$

$$f_{k|k-1}(\mathrm{x}_k\,|\,\mathrm{x}_{k-1})=\mathrm{N}(\mathrm{x}_k;\mathbf{F}_k\,\mathrm{x}_{k-1};Q_k) \tag{23}$$

$$g_k(\mathbf{y}_k|\mathbf{x}_k)=\mathrm{N}(\mathbf{y}_k;\mathbf{H}_k\,\mathrm{x}_k;\mathbf{R}_k) \tag{24}$$

The corresponding Gaussian PMC model can be expressed by the following formula [**8**]:

$$p(\xi_0)=\mathrm{N}\left(\xi_0;\begin{bmatrix}\mathrm{m}_0\\ \mathrm{H}_0\,\mathrm{m}_0\end{bmatrix};\begin{bmatrix}\mathrm{P}_0 & (\mathrm{H}_0\mathrm{P}_0)^T\\ \mathrm{H}_0\mathrm{P}_0 & \mathrm{R}_0+\mathrm{H}_0\mathrm{P}_0\mathrm{H}_0^T\end{bmatrix}\right) \tag{25}$$

$$p_{k|k-1}(\xi_k|\xi_{k-1})=\mathrm{N}(\xi_k;\mathbf{B}_k\xi_{k-1};\Sigma_k) \tag{26}$$

where $\xi_k=(\boldsymbol{x}_k,\boldsymbol{y}_k)$, and

$$\mathbf{B}_k=\begin{bmatrix}\mathbf{F}_k-\mathbf{F}_k^2\mathbf{H}_{k-1} & \mathbf{F}_k^2\\ \mathbf{H}_k\mathbf{F}_k-\mathbf{H}_k^2\mathbf{H}_{k-1} & \mathbf{H}_k^2\end{bmatrix},\Sigma_k=\begin{bmatrix}\Sigma_k^{11} & \Sigma_k^{21\,T}\\ \Sigma_k^{21} & \Sigma_k^{22}\end{bmatrix} \tag{27}$$

$$\Sigma_k^{11}=Q_k-\mathbf{F}_k^2\mathbf{R}_{k-1}(\mathbf{F}_k^2)^T \tag{28}$$

$$\Sigma_k^{21}=\mathbf{H}_kQ_k-\mathbf{H}_k^2\mathbf{R}_{k-1}(\mathbf{F}_k^2)^T \tag{29}$$

$$\Sigma_k^{22}=\mathbf{R}_k-\mathbf{H}_k^2\mathbf{R}_{k-1}(\mathbf{H}_k^2)^T+\mathbf{H}_kQ_k\left(\mathbf{H}_k\right)^T \tag{30}$$

## 4.2 Performance analysis

Let us now analyze quantitatively the tracking performance of PF-PMC-PHD filter and PF-HMC-PHD filter in a nonlinear system based on PMC model. We compute at each time step the OSPA metric and the target number estimation. Set

$$\mathbf{F}_k = \begin{bmatrix} 1 & \frac{\sin\Omega T}{\Omega} & 0 & -\frac{1-\cos\Omega T}{\Omega} \\ 0 & \cos\Omega T & 0 & -\sin\Omega T \\ 0 & \frac{1-\cos\Omega T}{\Omega} & 1 & \frac{\sin\Omega T}{\Omega} \\ 0 & \sin\Omega T & 0 & \cos\Omega T \end{bmatrix}, \mathbf{F}_k^2 = \begin{bmatrix} 0.7 & 0 \\ 0 & 0 \\ 0 & 0.7 \\ 0 & 0 \end{bmatrix} \tag{31}$$

$$\mathbf{H}_k = \begin{bmatrix} 1 & 0 & 0 & 0 \\ 0 & 0 & 1 & 0 \end{bmatrix}, \mathbf{H}_k^2 = \begin{bmatrix} 0.1 & 0 \\ 0 & 0.1 \end{bmatrix} \tag{32}$$

$$Q_k = \begin{bmatrix} 100 & 1 & 0 & 0 \\ 1 & 10 & 0 & 0 \\ 0 & 0 & 100 & 1 \\ 0 & 0 & 1 & 10 \end{bmatrix}, \quad R_k = \begin{bmatrix} 25 & 0 \\ 0 & 25 \end{bmatrix} \tag{33}$$

We generate uniformly a mean of 10 clutter measurements on region $V = [-2000, 2000] \times [-2000, 2000]$ with sampling period $T = 1s$, and simulation experiment step $N = 50$. There are 4 targets: target 1 and target 2appear at time $k = 1$, target 3 and target 4 appear at time $k = 20$. We track the position and velocity of targets in Cartesian coordinates, $\mathbf{x}_k = [p_{\mathrm{x},k}, \dot{p}_{\mathrm{x},k}, p_{\mathrm{y},k}, \dot{p}_{\mathrm{y},k}]^T$. We set $p_{s,k} = 0.98$, $p_{d,k} = 0.9$, and select $L = 2000$ for particle number of one target. Target trajectories and measurements of such a scenario is displayed in Figure.1.

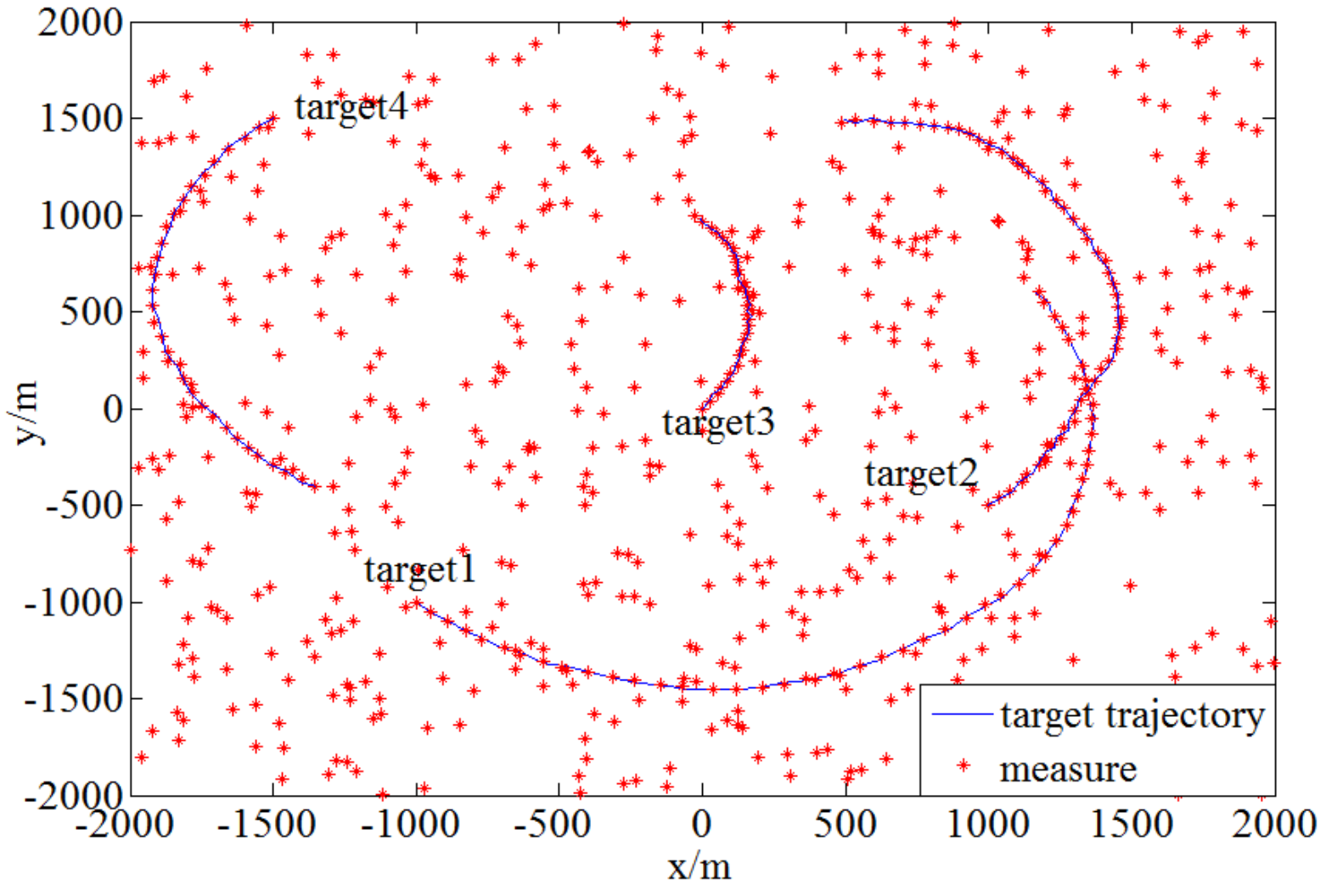

**Figure. 1** target trajectories and measurements

Figure. 2 shows the Tracking result of PF-PMC-PHD. Tracking result shows that the proposed PF-PMC-PHD filter can effectively achieve the target tracking.

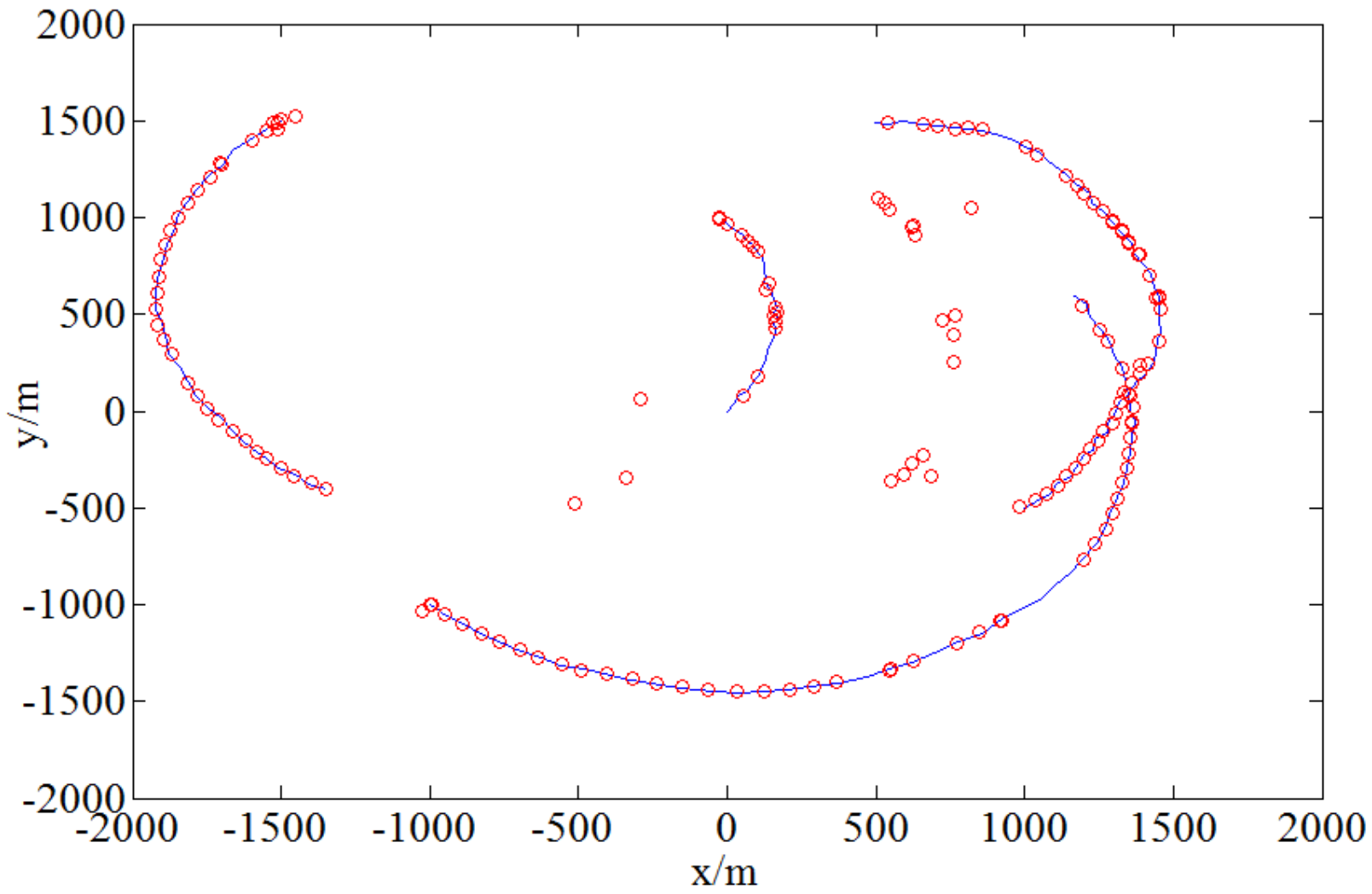

**Figure. 2** Tracking result of PF-PMC-PHD

Figure. 3 and Figure. 4 respectively show target number estimations and the OSPA metrics of PF-PMC-PHD filter, PF-HMC-PHD filter and UPF-PMC-PHD filter. Simulation result shows that the tracking performance of PF-PMC-PHD filter is better than that of PF-HMC-PHD filter although two filters share the same $p(\boldsymbol{x}_k|\boldsymbol{x}_{k-1})$ and $p(\boldsymbol{y}_k|\boldsymbol{x}_k)$. This is because that HMC model does not take into account the information given by the observation $\boldsymbol{y}_{k-1}$ in the case of given $\boldsymbol{x}_{k-1}$ and $\boldsymbol{y}_{k-1}$, and the uncertainty of the state $\boldsymbol{x}_k$ increases.The target number estimation error of PF-HMC-PHD filter is large, and the situation of missing target is serious. Simulation result also shows that UPF-PMC-PHD filter has higher accuracy of target number estimation and position error estimation than PF-PMC-PHD filter, because UPF-PMC-PHD filter uses UKF to generate the importance probability density function which makes full use of the new measurements, which results in higher accuracy of particle sampling and improved tracking performance.

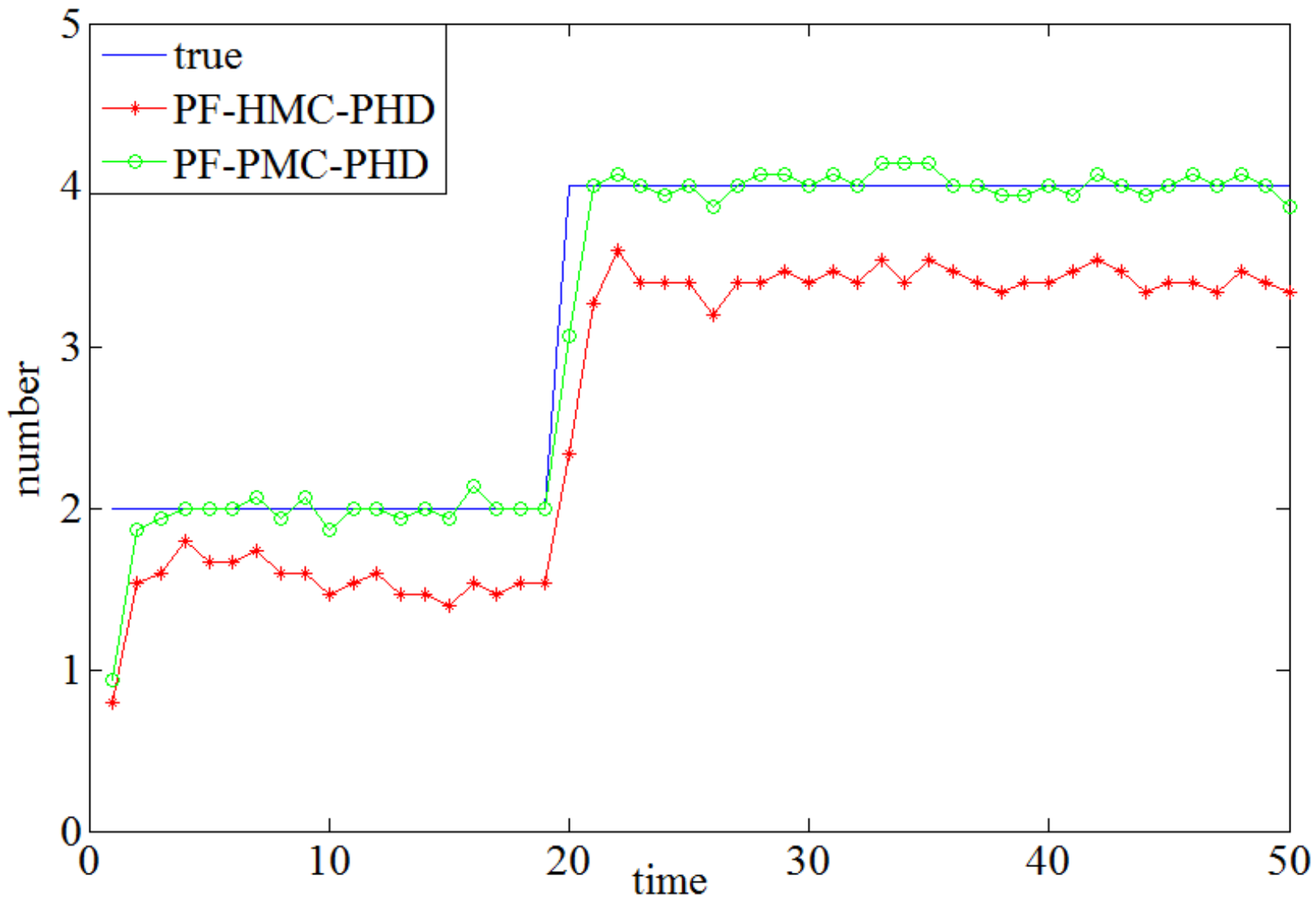

**Figure. 3** Target number estimations of PF-PMC-PHD and PF-HMC-PHD

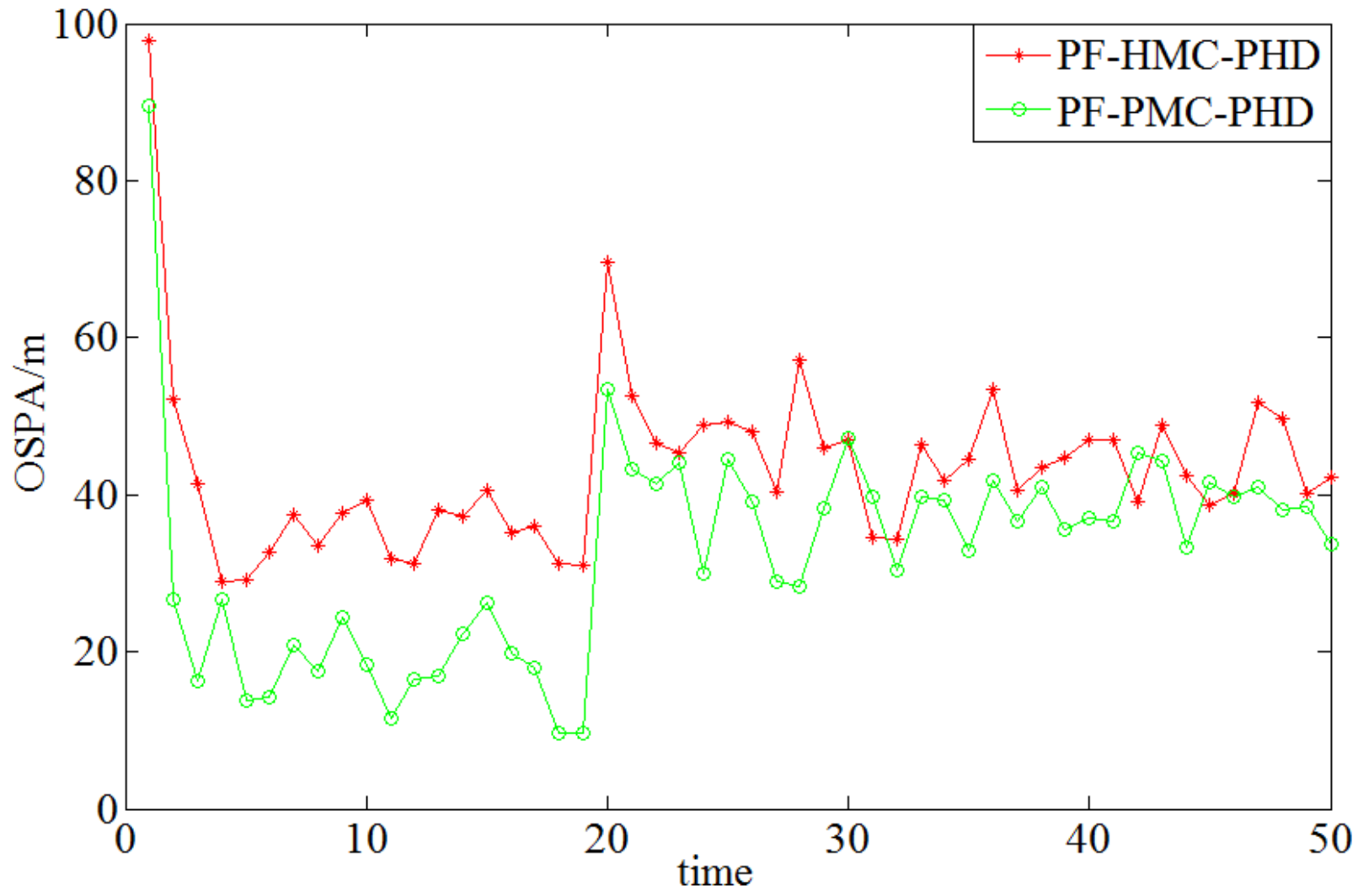

**Figure. 4** OSPA  metrics of PF-PMC-PHD and PF-HMC-PHD

## 5 Conclusions

The Pairwise Markov Chain (PMC) model is more universally suitable than the traditional Hidden Markov Chain(HMC) model. A PF-PMC-PHD filter is proposed for nonlinear multi-target tracking system based on PMC model. Simulation result proves the effectiveness of  the filter, but the tracking accuracy needs to be improved. Simulation result also shows that the performance of  PF-PMC-PHD filter is better than the traditional PF-HMC-PHD filter in a scenario where we kept the local physical properties of nonlinear  and Gaussian HMC models while relaxing their independence assumption.

## Declaration of Conflicting Interests

The author(s) declared no potential conflicts of interest with respect to the research, authorship, and/or publication of this article. analysis.

## Funding

This work was supported by the Natural Science Foundation of China under Grants 61141009.

## References

[1] Mahler R. An Introduction to Multisource-Multitarget Statistics and Its Applications. Lockheed Martin, Eagan, MN, Technical Monograph, 2000.

[2] Mahler R. Statistical Multisource-Multitarget Information Fusion. Norwood,MA: Artech House, 2007.

[3] Mahler R. Multitarget Bayes Filtering via First-Order Multitarget Moments. *IEEE Transactions on Aerospace and Electronic Systems*, 2003, 39(4): 1152-1178.

[4] Vo B N, Singh S, and Doucet A. Sequential Monte Carlo methods for multi-target filtering with random finite sets. *IEEE Transactions on Aerospace and Electronic Systems*. 2005,41(4): 1224-1245.

[5] GAO Yiyue, JIANG Defu, and LIU Ming. Particle-gating SMC-PHD Filter. *Signal Processing*, 2017, 130: 64 – 73.

[6] Vo B N and Ma W K. The Gaussian mixture probability hypothesis density filter. *IEEE Transactions on Signal Processing* , 2006, 54(11): 4091 - 4104.

[7] YANG Jinlong, LI Peng, YANG Le, et al.. An Improved ET-GM-PHD Filter for Multiple Closely-spaced Extended Target Tracking. *International Journal of Control, Automation and Systems*, 2017, 15(1): 468-472.

[8] Pieczynski W. Pairwise Markov chains and Bayesian unsupervised fusion. International International Conference on Information Fusion, Paris, France, 2000, MoD4: 24-31.

[9] Derrode S and Pieczynski W. Signal and image segmentation using pairwise Markov chains. *IEEE Transactions on Signal Processing*, 2004, 52(9):2477 – 89.

[10] Pieczynski W. Pairwise Markov chains. *IEEE Transactions on Pattern Analysis and Machine Intelligence*, 2003, 25(5): 634-639.

[11] Petetin Y and Desbouvries F. Bayesian multi-object filtering for pairwise Markovchains. *IEEE Transactions on Signal Processing*, 2013, 61(18): 4481-4490.

[12] Mahler R. Tracking targets with pairwise-Markov dynamics. International Conference on Information Fusion, Washington, D.C., 2015: 280-286.

[13] Mahler R, Consultant, Eagan, et al.. On multitarget pairwise-Markov models. *Society of Photo-Optical Instrumentation Engineers*. 2015, 9474 , 94740D: 1-12.

[14] Julier S J, Uhlmann J K, and Durrant-Whyte H F. A new method for nonlinear transformation of means and covariances in filters and estimators. *IEEE Transactions on Automatic Control*, 2000, 45(3): 477–482.

[15] Julier S J, and Uhlmann J K. Unscented filtering and nonlinear estimation. *IEEE Transactions on Signal Processing*, 2004, 92(3): 401-422.